\documentclass[Aps,prl,reprint,twocolumn,showpacs,groupedaddress]{revtex4-2}
\usepackage{amssymb}
\usepackage{float}
\usepackage{graphicx}
\usepackage{color}
\usepackage{soul}
\usepackage[pdftex,colorlinks=true,linkcolor=red,citecolor=blue]{hyperref}

\begin{document}

\title[]{Electronic correlations and topology in Kondo insulator PuB$_6$}

\author{K. Gofryk$^{1,2}$}
\email{gofryk@inl.gov}%
\author{S. Zhou$^2$}
\author{N. Poudel$^3$}
\author{N. Dice$^3$}
\author{D. Murray$^3$}
\author{T. Pavlov$^3$}
\author{C. Marianetti$^4$}

\affiliation{$^1$Center for Quantum Actinide Science and Technology, Idaho National Laboratory, Idaho Falls, Idaho 83415, USA}
\affiliation{$^2$Glenn T. Seaborg Institute, Idaho National Laboratory, Idaho Falls, Idaho 83415, USA}
\affiliation{$^3$Idaho National Laboratory, Idaho Falls, Idaho 83415, USA}
\affiliation{$^4$Department of Applied Physics and Applied Mathematics, Columbia University, New York, New York 10027, USA}

\date{\today}%

\begin{abstract}
Utilizing a combination of dynamical mean field theory and density functional theory (DMFT/DFT), it has been theoretically proposed that PuB$_6$ is a strongly correlated topological insulator characterized by nontrivial $\mathbf{Z}_{2}$ topological invariants and metallic surface states (\textit{X. Deng et al., Phys. Rev. Lett. 111, 176404 (2013)}). Here, we demonstrate through low-temperature magneto-transport measurements and first-principles calculations that PuB$6$ exhibits characteristics of a topological Kondo insulating state. These features include a transition in electrical resistivity from high-temperature, thermally activated behavior with a narrow gap at the Fermi level ($\Delta{\rho} \sim$ 20 meV) to a distinctive low-temperature plateau, as well as a surface-to-volume dependence of electrical resistivity at low temperatures. The topological nature of PuB$_6$ is further supported by the theoretical calculations, which show that GGA+$U$ is capable of capturing electronic, topological, and lattice properties of PuB$_6$ with much lower computational cost than DMFT. 
\end{abstract}


\maketitle


\textit{Introduction} - The concept of strongly correlated topological insulators is highly appealing not only because the surface states, protected from back-scattering by time-reversal symmetry, may host massless charge carriers with locked helical spin polarization, but also because the surface of such correlated system may exhibit non-trivial electronic structures not present in conventional band insulators \cite{1,2}. The 5$f$-electron materials, in particular, possess all the essential ingredients for hosting novel topological phenomena. These include strong spin-orbit coupling, electrons of opposite parity (e.g., 5$f$ vs. 5$d$) that are inverted in energy in particular sections in the Brillouin zone, and a large $f$-$f$ overlap, that results in larger energy scales. Since most relevant electronic interactions have similar energy scales, combining strong electronic interactions and topology offers a unique way to control topological properties. To date, only a handful 5$f$-electron-based systems have been predicted or demonstrated to host topological phenomena, that include new family of topological Mott insulators, $AnX$, where $An$ = Pu, Am and $X$ = pnictogen and chalcogen elements \cite{claudia}, U$_{3}$Bi$_{4}$Ni$_{3}$ Kondo insulator \cite{ran}, UCo$_{0.8}$Ru$_{0.2}$Al ferromagnet (Weyl semimetal with large Nernst effect) \cite{asaba}, UTe$_{2}$ (Weyl superconductor) \cite{UTE2a,UTE2b}, UOTe antiferromagnet (Dirac semimetal)\cite{UOTe} or/and PuB$_{4}$ topological insulator \cite{PuB4,PuB4b}. 

Topological Kondo insulators (TKIs) in particular, have garnered significant interest in the field of strongly correlated electronic systems, with special attention devoted to samarium hexaboride (SmB$_6$)\cite{3}, which has been proposed as the first member of this family \cite{4,5,6, 9,9a}. This discovery has spurred huge scientific interests, both theoretically\cite{9,9a,10,11,12} and experimentally \cite{13,14,15,16,17,19,20}, although several inconsistencies remain regarding the specifics of its topological nature. Similar CeB$_6$ and YbB$_6$ have also been predicted to exhibit non-trivial electronic structure \cite{CeB6T,YbB6T}, but experimental studies have not supported these predictions \cite{CeB6E,YbB6E}. Recently, it has been proposed theoretically that PuB$_6$ is an intermediate valent, strong topological insulator with nontrivial $\textbf{Z}_2$ topological invariants \cite{PuB6}. Its surface states are predicted to contain three Dirac cones with a large Fermi pocket at the $X$ point, a characteristic feature of cubic topological Kondo insulators \cite{12}. However, experimental validation of the electronic ground state and its relationship to the topological characteristics in PuB$_6$ remain lacking.

\begin{figure}[t]
\centering
\includegraphics[width=0.85\linewidth]{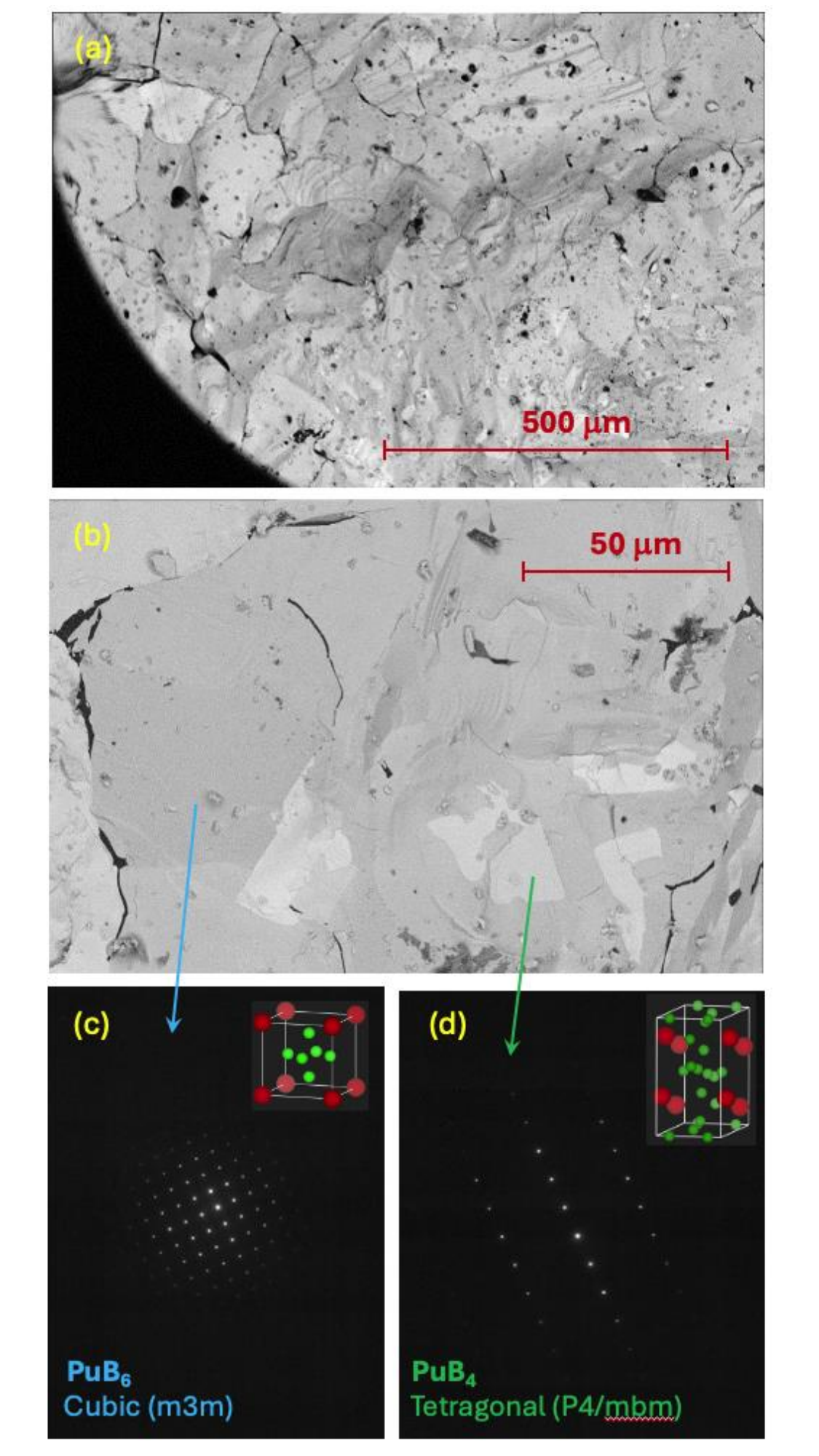}
\caption{(a,b) Scanning electron microscope micrographs of arc-melted sample of PuB$_6$ showing a mixture of PuB$_6$ and PuB$_4$, as identified by selected area diffraction (c,d). The insets show the crystallographic structure of cubic (m3m) PuB$_{6}$ and tetragonal (P4/mbm) PuB$_{4}$ (see more details in the text).}
\label{Fig1}
\end{figure}

In this Letter, we address this topic by investigating the low-temperature electronic properties of PuB$_6$ through synthesis and micromachining characterization, magnetotransport measurements, and detailed electronic structure calculations. The electrical resistivity reveals a narrow energy gap at the Fermi level and can be described by a two-conductivity channels model that accounts for both metallic and semiconducting components, similar to other topological insulators. The presence of surface states is further supported by the low-temperature resistivity plateau and the dependence of electrical resistivity on crystal geometry (surface-to-volume ratio). The magnetoresistivity measurements indicate incomplete surface dominance of the interplay between Dirac and conventional carrier rather than a single pristine topological transport channel in PuB$_6$. This work demonstrates that PuB$_6$ is a compelling candidate for investigating the influence of electronic correlations and topology, supporting theoretical predictions of an insulating bulk with metallic surface states. We discuss the implications of this study and the role of increased $f$-$f$ overlap and the resulting energy scales on the topological characteristics when transitioning from widely studied 4$f$-electron systems to actinide-based topological materials.

\textit{Methods} - Polycrystalline samples of PuB$_6$ were synthesized by a standard arc melting technique. In PuB$_6$, this synthesis method can lead to a multi-phase sample that contains a mixture of cubic PuB$_6$ and tetragonal PuB$_4$, \cite{synt,SM1} rendering it unsuitable for bulk measurements. To overcome these issues, we utilized an FEI Helios plasma Focus Ion Beam (PFIB) microscope to characterize, extract, and prepare micro-sized crystals of PuB$_6$ that are suitable for low temperature magnetotransport measurements (see also Ref.\onlinecite{kappa}). The FIB micromachining approach was proven revolutionary for studying the magneto-transport properties of various $p,d,f$-electron topological materials \cite{f1,f2,kappa} and very recently uranium systems \cite{FIBUa,FIBUb}. The polycrystalline sample was first examined in the PFIB microscope using back-scattered electron imaging, energy dispersive spectroscopy, and electron back-scattered diffraction to locate a single crystal grain of interest and verify its crystallographic structure and grain orientation (see Fig.~\ref{Fig1}). Our polycrystalline button was a mixture of PuB$_6$ and PuB$_4$, and the structural analysis confirmed that the expected cubic (Pm3m) and tetragonal (P4/mbm) crystal structures can be identified with lattice parameters: $a$ = 4.129(3)~\AA,~and $a$ = $b$ = 7.313(8)~\AA~and $c$ = 4.098(9)~\AA, respectively (see Fig.~\ref{Fig1}). These values are very close to the ones previously reported for these phases \cite{synt,B1,B2,B3}. It is worth mentioning, that due to the multi-phase nature of the samples obtained we were unable to perform any bulk measurements such as heat capacity, magnetization, and especially angle-resolved photoemission spectroscopy to directly probe electronic structure in PuB$_{6}$. Once a suitable grain of PuB$_6$ phase was located, extracted, and structurally characterized (crystallographic orientation) it was transferred to a sapphire chip with contacts patterned by photolithography. Electrical contacts (platinum in our case) were then deposited onto the crystal with ion-assisted chemical vapor deposition.\cite{kappa,hosen} Further details regarding the preparation and quantity of electrical contacts can be found in Ref.\cite{SM1}. Low-temperature magnetotransport measurements were performed using a Quantum Design DynaCool-9 system and a standard four-point method. DFT calculations were carried out using the projector augmented-wave (PAW) method \cite{23,24}, as implemented in the Vienna ab initio Simulation Package (VASP) \cite{25,26}. The generalized gradient approximation (GGA) as formulated by Perdew, Burke, and Ernzerhof (PBE) \cite{27} was employed with a plane-wave cutoff energy of 600 eV. The $14\times14\times14$ $\Gamma$-centered k-point mesh and an energy convergence criterion of $10^{-6}$ eV were applied for the primitive cell. DFT+$U$  was applied on Pu 5\emph{f} electrons by using the simplified rotationally invariant approach \cite{22,28}. All symmetry was turned off, and spin-orbit coupling (SOC) was applied except when explicitly mentioned. Furthermore, the occupation matrix control approach was used to address the issue of many metastable electronic states in DFT+$U$ \cite{29}, while more details can be found in the Supplemental Materials \cite{SM2}. Finally, to match with the nonmagnetic (NM) semiconducting nature of PuB$_6$ from DMFT calculations \cite{PuB6} and experiments, we employed DFT calculations in NM order. However, please note that by using GGA+$U$ for PuB$_6$, magnetic orderings are predicted to have lower energy than NM ordering, while having a metallic nature (see the Supplementary Materials \cite{SM2}).  This is due to the limitations of DFT+$U$, where it predominantly captures singlet states without symmetry breaking. Similar results have been reported in $\delta$-Pu \cite{41,42}.    

\begin{figure}[t]
\centering
\includegraphics[width=0.9\linewidth]{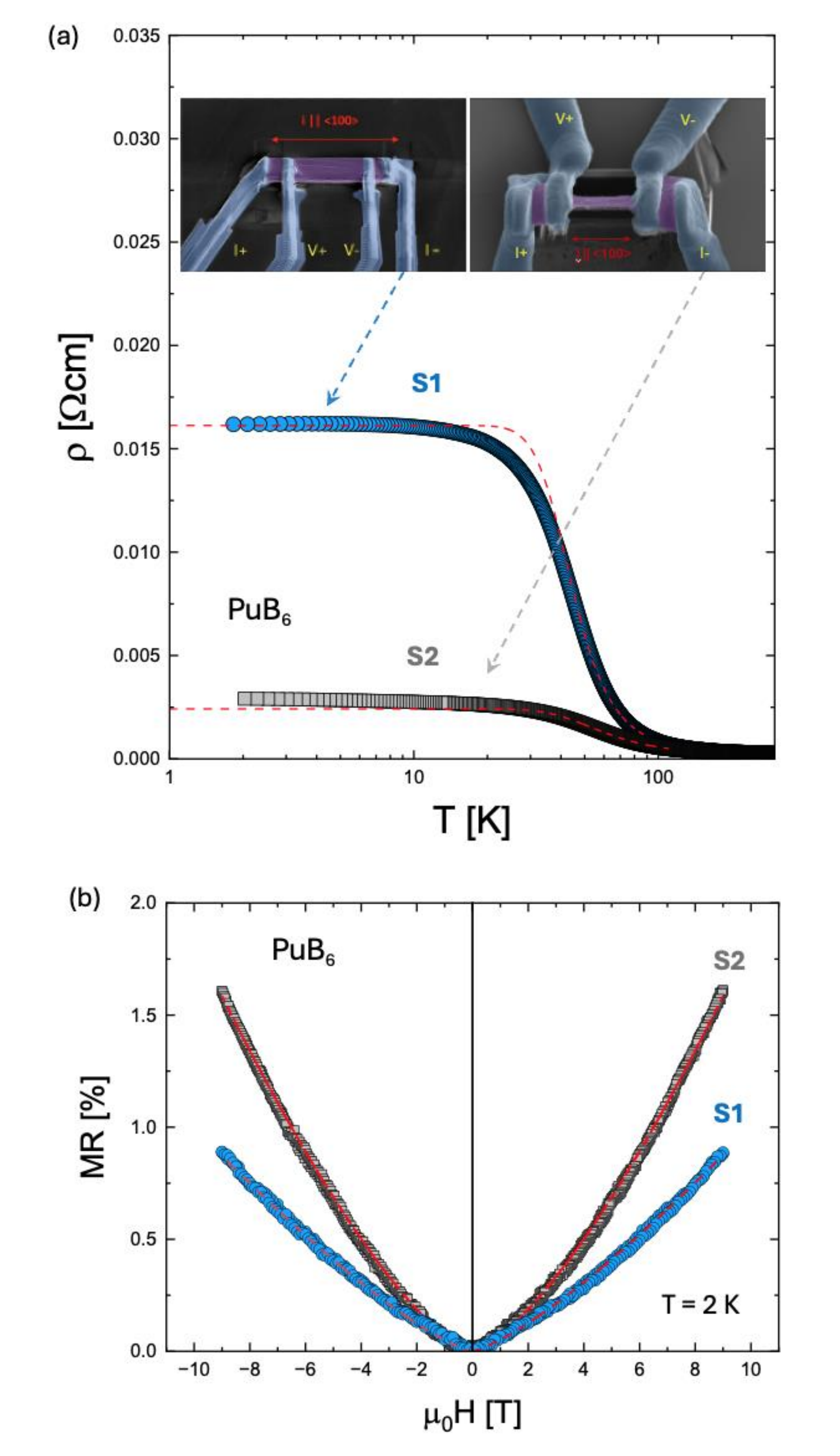}
\caption{(a) The temperature dependence of the electrical resistivity of the PuB$_{6}$ microcrystals. Samples dimensions (thickness $\times$ width $\times$ length) are 3.4 $\times$ 6.7 $\times$ 31 $\mu m^3$ and 0.5 $\times$ 1 $\times$ 20 $\mu m^3$, respectively for samples S1 and S2. The dashed lines represent fits to the data using a two-channel conductance model (see the text). The insets show PFIB lamellas of PuB$_{6}$ prepared for low-temperature transport measurements ($i || <100>$). (b) The magnetic field dependence of transverse magnetoresistivity of PuB$_{6}$ crystals measured at T = 2 K. The dashed line represents the relation MR $\sim$ $H^{1.3}$ for sample S1 and MR $\sim$ $H^{1.4}$ for sample S2.}
\label{Fig2}
\end{figure}

\textit{Magnetotransport studies} - Figure \ref{Fig2}a displays the temperature dependence of the electrical resistivity of a single-crystal lamellas of PuB$_6$, measured for crystals with different geometries and with the electrical current ($i$) applied along the $<$100$>$ crystallographic direction (see the insets of Figure~\ref{Fig2}). As shown, the $\rho(T)$ curves exhibit semiconducting-like behavior down to $\sim$30 K, below which a resistivity plateau is observed. In this regard, the overall $\rho(T)$ dependence resembles that observed in non-magnetic topological insulators, where bands inversion leads to the existence of the metallic surface modes protected by time-reversal symmetry\cite{top1,top2,top3}. The transition from high-temperature, thermally activated behavior to a low-temperature plateau in $\rho(T)$ is interpreted as a shift from bulk state-dominated conduction to surface state-dominated conduction. This scenario is also consistent with the surface-to-bulk ratio dependence of the resistivity. As shown in Figure \ref{Fig2}a, the resistivity decreases with increasing surface-to-bulk ratio ($S/V$), with $S/V$ values of 0.9 and 6 $\mu$m$^{-1}$ for samples S1 and S2, respectively, as expected due to the reduction in overall bulk conductance. To account for both surface and bulk contributions to the resistivity of PuB$_6$ crystal, we analyzed the electrical resistivity data using a two-channel model (see the red dashed line in Figure~\ref{Fig2}a), which has previously been used to describe $\rho(T)$ dependence of SmB$_6$ \cite{top3}. In this approach, the total electrical conductivity $\sigma(T) = \rho(T)^{-1}$ is given by the equation: $\sigma_{tot}(T) = \sigma_{s}+\sigma_{b}e^{-\frac{\Delta_{\rho}}{k_{B}T}}$, where $\sigma_{s}$ is the surface contribution to the electrical conductivity (independent of temperature), $\sigma_{b}$ is the bulk contribution to the conductivity, $\Delta_{\rho}$ is the energy gap at the Fermi level, and $k_{B}$ is the Boltzmann constant. The analysis yields: $\Delta_{\rho}$ = 20 and 18 meV, $\sigma_{s}$ = 62.1 and 405.4 $\Omega$m$^{-1}$, $\sigma_{b}$= 13.9$\times$10$^{-3}$ and 11.9$\times$10$^{-3}$ $\Omega$m$^{-1}$, for S1 and S2 samples, respectively. As expected for topological insulators, the obtained energy gap is similar for the two samples with different surface-to-volume ratio. However, the values or $\sigma_{s}$ and $\sigma_{b}$ show a clear contrast in their relationship with samples' surface-to-volume ratio. While $\sigma_{b}$ parameter is relatively independent of the samples' $S/V$ ratio, $\sigma_{s}$ exhibits a clear dependence on the samples geometry with $\sigma_{s}$ being much larger for S2, in agreement to what is expected for topological insulators. Furthermore, the value of $\Delta_{\rho}$ is 5-6 times smaller than $\Delta_{DFT} \sim$ 100-120 meV obtained by DFT calculations (see below). Interestingly, PuB$_4$ has also been predicted to host a topological insulating ground state with an energy gap of about 250 meV due to strong spin-orbit interactions \cite{PuB4,PuB4b}. The derived resistivity gap has been estimated to be 35 meV \cite{PuB4}. 

\begin{figure}[t]
\centering
\includegraphics[width=0.8\linewidth]{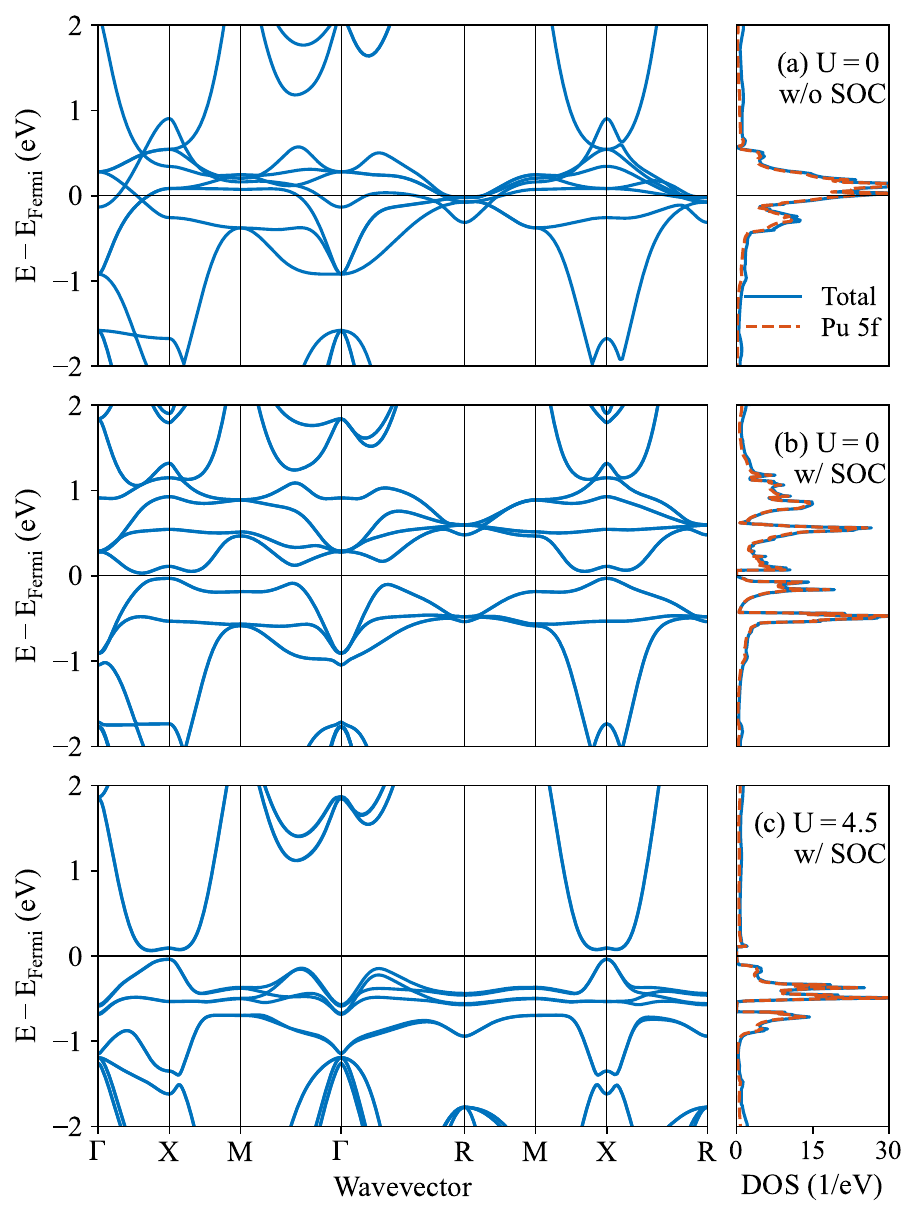}
\caption{Calculated electronic bands and DOS by using (a) GGA ($U=0$ and without SOC), (b) GGA+SOC ($U=0$), and (c) GGA+$U$+SOC ($U=4.5$ eV). The solid blue curves represent total DOS, while dashed red curves present the partial DOS of Pu 5$f$ electrons.}
\label{fig_bands}
\end{figure}

Figure~\ref{Fig2}b shows the magnetic field dependence of the transverse magnetoresistivity (MR) measured at 2 K. As can be seen, the MR(H) curves can be described by the form MR $\propto$ H$^{1.3}$ and H$^{1.4}$ respectively for samples S1 and S2 (see red dashed lines in Fig.~\ref{Fig2}b). The classical magnetoresistance in metals or doped semiconductors with a closed free-electron Fermi surface increases quadratically with increasing magnetic field $H$ for $\mu_{H}<$ 1 and saturates when $\mu_{H}>$ 1 ($\mu$ is the zero-magnetic-field mobility). Magnetoresistivity in ideal topological insulators is often discussed in the context of linear field dependence (MR $\propto H$) at low temperatures, especially for surface states with Dirac-like dispersion in the quantum limit. Such behavior has previously been observed in surfaces of topological insulators MnBi$_{2}$Te$_{4}$ \cite{LMR1}, Ru$_{2}$Sn$_{3}$ \cite{LMR2}, Dirac semimetals Cd$_{3}$As$_{2}$ \cite{LMR4}, and other 3D Dirac materials \cite{LMR5,LMR6}. A weak antilocalization effects and the linear magnetoresistivity was observed in the orbital component of MR at $T\sim$50-100 mK in SmB$_{6}$.\cite{LMR3} In topological insulators with both bulk and surface conduction, quadratic bulk MR from parabolic bands and linear MR from the Dirac-like surface states can combine into as intermediate power-law dependence\cite{int}, as observed in PuB$_{6}$. Additionally, as can be seen in Fig.~\ref{Fig2}b, sample S1 (with more bulk weight) exhibits lower MR, while sample S2 (with more surface dominance) show higher MR. Further magnetotransport experiments, particularly at very low temperatures, will be essential to draw definitive conclusions about the role of potential disorder, in-gap states and/or localization effects in PuB$_6$.

\begin{figure}[t]
\centering
\includegraphics[width=0.85\linewidth]{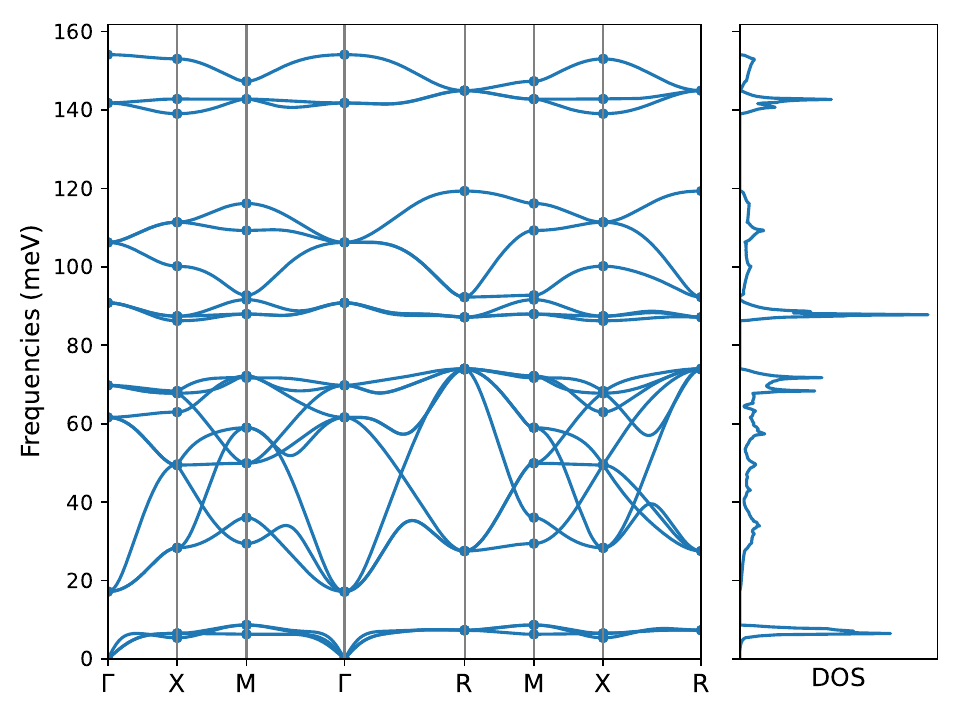}
\caption{Calculated phonon dispersion and density of states (DOS) of PuB$_6$ by using GGA+U+SOC ($U=4.5$ eV). The solid points were directly computed, while the corresponding lines are Fourier interpolations.}
\label{phonon}
\end{figure}

\textit{DFT calculations} - Here DFT+$U$ calculations are performed for PuB$_{6}$ as a static description of electron interactions at $T=0$ K. Compared with DFT+DMFT, our DFT+$U$ results show good agreements of predicted electronic properties with previous studies \cite{PuB6,PuBx}, while only taking a fraction of the computational cost. Therefore, DFT+$U$ can be applied to study some complex properties, e.g., phonon properties, in PuB$_6$. The effect of SOC and the Hubbard $U$ on the calculated electronic bands and density of states (DOS) is presented in Figure~\ref{fig_bands}. In GGA ($U=0$ and without SOC), the Pu 5$f$ electrons result in several nearly flat bands near the Fermi level $E_{Fermi}$, producing a metallic nature. By applying SOC, a band gap of 60.7 meV is opened. The band inversion produced by SOC, which is a signature of topological band insulator, can be observed in PuB$_6$ near the $X$ point. Furthermore, by adding the Hubbard $U$, the Pu 5$f$ states are further localized. Typically, the Hubbard $U$ for Pu is in the range of $4$-$4.5$ eV \cite{31,32,33}. For PuB$_6$, we used the linear response approach \cite{35} to determine the Hubbard $U$, yielding $U=3.74$ eV (see \cite{SM5}), which is reasonably close to the $4$-$4.5$ eV range. While the Hubbard $U$ value can quantitatively affect the value of lattice parameters and the band gaps, a band gap produced by band inversion is always observed for $U < 6.5$ eV (see the Supplementary Materials \cite{SM5}). Notably, our computed band structure shows great agreement with Ref.\cite{PuB6}, both for comparing our GGA results (Fig.~\ref{fig_bands} panel (b)) to their LDA results, and our GGA+$U$ (panel (c)) to their LDA+DMFT results, which uses the same $U=4.5$ eV. We further developed a tight-binding model based on GGA+$U$+SOC calculation, and computed the $Z_2$ topological invariant as well as electronic bands on (100) surface (see computational details and results in \cite{SM5}). The computed $Z_2$ number is (1,111), with a conducting surface band at $\bar{X}$ point, both of which are in good agreement with Ref.\cite{PuB6}. Additionally, we compute the phonon dispersion of PuB$_6$ by using GGA+$U$ (see Figure~\ref{phonon}) and the lone irreducible derivative approach \cite{36} with the 2$\times$2$\times$2 supercell. While the absence of imaginary phonon frequencies confirms the structural stability in our calculation, and the phonon characteristics also indicate the d-f electron hybridization in PuB$_6$: similar to SmB$_6$ \cite{37}, phonon softening is observed for the two lowest optical branches near the $\Gamma$ point and the longitudinal acoustic branch near the $X$ point. Furthermore, phonon interaction calculations facilitated by DFT+$U$ present new opportunities to explore the insulating characteristics of PuB$_6$, for examples, by isolating the contributions of electrons and phonons to thermal transport or illustrating the impact of vacancies, as observed in SmB$_6$ \cite{38,39,40}.

\textit{Summary and Outlook} - In summary, we performed synthesis, PFIB structural characterizations and micromachining, and low-temperature magnetotransport measurements, along with detailed DFT calculations of PuB$_{6}$. We demonstrate that PFIB micromachining and characterization serve as excellent tools for investigating the magnetotransport properties of this 5$f$-electron quantum material. This is particularly crucial for transuranic materials, where conventional synthesis methods might lead to multiphase samples. Based on the results obtained, we provide experimental evidence that PuB$_6$ exhibits characteristics of topological Kondo insulating state such as the presence of a narrow gap at the Fermi level ($\Delta_{\text{DFT}}$ $\sim$ 100 meV, $\Delta_{\rho}$ $\sim$ 20 meV), a characteristic low-temperature plateau of the electrical resistivity, the resistivity dependence on samples' surface-to-volume ratio, and MR dependence characteristic of topological insulators with both bulk and surface conduction. The intricate details of the non-trivial electronic structure of PuB$_6$ are further supported by theoretical calculations, where we show that GGA+$U$ can accurately capture electronic, topological, and lattice properties of PuB$_6$ with much lower computational cost than DMFT. These new results not only expand our understanding of this system beyond recent band structure calculations but also pave the way for a more profound comprehension between electronic correlations and topology in 5$f$-electron materials.

\textit{Acknowledgments} - Authors would like to thank Jason Jeffries and Mitchel Meyer for fruitful discussions. Work supported by the US Department of Energy, Basic Energy Sciences, Materials Sciences, and Engineering Division.

All data generated in this study are available from the authors upon request.


\begin{thebibliography}{99}

\bibitem{1} For a review, see, e.g., M.Z. Hasan and C.L. Kane, Rev. Mod. Phys. 82, 3045 (2010).
\bibitem{2} For a review, see, e.g., N.P. Armitage, E.J. Mele, and A. Vishwanath, Rev. Mod. Phys. 90, 015001 (2018).

\bibitem{claudia} X. Zhang et al., Science, 335,1464 (2012).

\bibitem{ran} C. Broyles et al., Sci. Adv. 11, eadq9952 (2025)

\bibitem{asaba} T. Asaba et al., Sci. Adv. 7, eabf1467 (2021)

\bibitem{UTE2a} Lin Jiao et al., Nature 579, 523 (2020)
\bibitem{UTE2b} T. Shishidou et al., Phys. Rev. B 103, 104504 (2021)

\bibitem{UOTe} C. Broyles et al., Adv. Mater. 37, 2414966 (2025)
\bibitem{PuB4} H. Choi et al., Phys. Rev. B 97, 201114 (2018).
\bibitem{PuB4b} Dong-Choon Ryu et al., J. Am. Chem. Soc. 142, 19278 (2020)

\bibitem{3} A. Menth, E. Buehler, and T. H. Geballe, Phys. Rev. Lett. 22, 295 (1969).
\bibitem{4} G. Aeppli and Z. Fisk, Comments Condens. Matter Phys. 16, 155 (1992). 
\bibitem{5} H. Tsunetsugu, M. Sigrist, and K. Ueda, Rev. Mod. Phys. 69, 809 (1997).
\bibitem{6} P. Riseborough, Adv. Phys. 49, 257 (2000).

\bibitem{9} M. Dzero, K. Sun, V. Galitski, and P. Coleman, Phys. Rev. Lett. 104, 106408 (2010).

\bibitem{9a} M. Dzero, J. Xia, V. Galitski, and P. Coleman, Annu. Rev. Condens. Matter Phys. 7, 249 (2016).

\bibitem{10} M. Dzero, K. Sun, P. Coleman, and V. Galitski, Phys. Rev. B 85, 045130 (2012).

\bibitem{11} F. Lu, J. Zhao, H. Weng, Z. Fang, and X. Dai, Phys. Rev. Lett. 110, 096401 (2013). 

\bibitem{12} V. Alexandrov, M. Dzero, P. Coleman, Phys. Rev. Lett. 111, 226403, (2013)

\bibitem{13} H. Miyazaki, T. Hajiri, T. Ito, S. Kunii, and S.-i. Kimura, Phys. Rev. B 86, 075105 (2012).

\bibitem{14} X. Zhang et al., Phys. Rev. X 3, 011011 (2013).

\bibitem{15} D.J. Kim et al., Scientific Reports 3, 3150 (2013)

\bibitem{16} M. Neupane et al., Nature Communications 4, 2991 (2013) 

\bibitem{17} N. Xu et al., Phys. Rev. B 88, 121102(R) (2013)

\bibitem{19} J. Jiang et al., Nature Communications 4, 3010 (2013) 

\bibitem{20} D. J. Kim, J. Xia, and Z. Fisk, Nature Materials 13, 466 (2014)

\bibitem{YbB6T} H. Weng et al., Phys. Rev. Lett. 112, 016403 (2014)

\bibitem{CeB6T} R. Zhang et al., Phys. Rev. B 105, 165140 (2022)

\bibitem{YbB6E} Chang-Jong Kang et al., Phys. Rev. Lett. 116, 116401 (2016)
\bibitem{CeB6E} M. Neupane et al., Phys. Rev. B 92, 104420 (2015)

\bibitem{PuB6} X. Deng, K. Haule, and G. Kotliar, Phys. Rev. Lett. 111, 176404 (2013)

\bibitem {synt} Harry A. Eick, Inorg. Chem. 4, 8, 1237–1239 (1965)

\bibitem{SM1} See the Supplemental Material [URL], Section VI: Synthesis and Characterization, which provides a more detailed description of the synthesis process, as well as the preparation and evaluation of the electrical contacts.

\bibitem {kappa} N. Poudel et al., Appl. Phys. Lett. 126, 192201 (2025)

\bibitem{f1} P.J.W. Moll et al., Nature Materials 9, 628 (2010); Science 351, 1061 (2016); Nature 535, 266 (2016)

\bibitem{f2} I. Antonyshyn et al., Angew. Chem. Int. Ed. 59, 11136 (2020).

\bibitem{FIBUa} S. Hamann et al., Phys. Rev. B 104, 155123 (2021)
\bibitem{FIBUb} T. Helm et al., Nature Comm. 15, 37 (2024)

\bibitem{B1} P. Rogl and P.E. Potter, Journal of Phase Equilibria 18, 467 (1997)

\bibitem{B2} B .J. McDonald and W.I. Stuart, Acta Crystallogr., 13, 447-448 (1960)

\bibitem{B3} A. B. Shick, L. Havela, A. I. Lichtenstein, and M. I. Katsnelson, Scientific Reports 5, 15429 (2015)

\bibitem{hosen} M. Hosen et al., Scientific Reports 10, 12961 (2020) 

\bibitem{23} P. E. Blöchl, PPhys. Rev. B 50, 17953 (1994).
\bibitem{24} G. Kresse and D. Joubert, Phys. Rev. B 59, 1758 (1999).
\bibitem{25} G. Kresse and J. Hafner, Phys. Rev. B 47, 558(1993).
\bibitem{26} G. Kresse and J. Furthmuller, Phys. Rev. B 54, 11169 (1996).
\bibitem{27} J. P. Perdew, K. Burke, and M. Ernzerhof, Phys. Rev. Lett. 77, 3865 (1996).
\bibitem{28} S. L. Dudarev et al., Physical Review B 57, 1505 (1998); Phil. Mag. B 75, 613 (1997)

\bibitem{22} V. I. Anisimov, J. Zaanen, and O. K. Andersen, Phys. Rev. B 44, 943 (1991)

\bibitem{29} B. Dorado et al., Journal of Physics: Condensed Matter 25, 333201 (2013)

\bibitem{SM2} See the Supplemental Material [URL], Sections I and II, which provide a more detailed description of the ground state search and magnetic ordering in PuB$_{6}$.

\bibitem{top1} W. Ko et al., Scientific Reports, 3, 2656 (2013) 

\bibitem{top2} M. Pickem, E. Maggio, and J.M. Tomczak, Communications Physics 4, 226 (2021) 

\bibitem{top3} P. Syers, D. Kim, M.S. Fuhrer, and J. Paglione, Phys. Rev. Lett. 114, 096601 (2015)


\bibitem{LMR1} X. Lei et al., Phys. Rev. B 102, 235431 (2020)

\bibitem{LMR2} Y. Shiomi and E. Saitoh, AIP Advances 7, 035011 (2017)

\bibitem{LMR4} J. Feng et al., Phys. Rev. B 92, 081306(R) (2015)

\bibitem{LMR5} M. Novak et al., Phys. Rev. B 91, 041203(R) (2015)
\bibitem{LMR6} S.K. Kushwaha et al., APL Mater. 3, 041504 (2015)

\bibitem{LMR3} S. Thomas et al., Phys. Rev. B 94, 205114 (2016)

\bibitem{int} S. Singh et al., J. Phys.: Condens. Matter 29, 505601 (2017)

\bibitem{PuBx} Haiyan Lu and Li Huang, J Phys.: Condens. Matter 34 215601 (2022)

\bibitem{31} J. H. Shim, K. Haule, and G. Kotliar, Nature (London) 446, 513 (2007).
\bibitem{32} M.-T. Suzuki and P. M. Oppeneer, Phys. Rev. B 80, 161103(R) (2009).
\bibitem{33} J.-X. Zhu et al., Europhys. Lett. 97, 57001 (2012); Nat. Commun. 4, 2644 (2013)
\bibitem{35} M. Cococcioni and S. de Gironcoli, Phys. Rev. B 71, 035105 (2005).

\bibitem{SM5} See the Supplemental Material [URL], Sections III-V, which provide a more detailed description on the effect of Hubbard U on the lattice structure and electronic band structure, Tight-Binding model used, and the linear response approach to determine the Hubbard U.

\bibitem{36} L. Fu, M. Kornbluth, Z. Cheng, and C. A. Marianetti, Physical Review B 100, 014303 (2019).

\bibitem{37} P.A. Alekseev et al., Europhysics Letters, 10, 457 (1989).

\bibitem{38} M-E. Boulanger et al., Phys. Rev. B 97, 245141 (2018).

\bibitem{39} J. Knolle and N. R. Cooper, Phys. Rev. Lett, 118, 096604 (2017).

\bibitem{40} M. E. Valentine et al., Phys. Rev. B 94, 075102 (2016).

\bibitem{41} Shick et al. Europhys. Lett. 69, 588 (2005).

\bibitem{42} Pourovskii
et al Physical Review B 75, 235107 (2007).

\end{thebibliography}
\end{document}